\documentclass[conference]{IEEEtran}
\IEEEoverridecommandlockouts
\usepackage{multirow}
\usepackage{cite}
\usepackage{amsmath,amssymb,amsfonts}
\usepackage{algorithmic}
\usepackage{graphicx}
\usepackage{textcomp}
\usepackage{array}
\usepackage{xcolor}
\usepackage{booktabs}
\usepackage{url}
\def\BibTeX{{\rm B\kern-.05em{\sc i\kern-.025em b}\kern-.08em
    T\kern-.1667em\lower.7ex\hbox{E}\kern-.125emX}}
\begin{document}

\title{TRACE-5G: Trust-Aware Runtime Attestation and Continuous Enforcement for 5G User Equipment}

\author{
\IEEEauthorblockN{
Al Nahian Bin Emran$^{*}$, Rajendra Paudyal$^{*}$, Rajendra Upadhyay$^{*}$, Arupjyoti Bhuyan$^{\dagger}$, Duminda Wijesekera$^{*}$
}
\IEEEauthorblockA{
$^{*}$Mason Innovation Labs, George Mason University, 3434 Washington Blvd Suite 1300, Arlington, VA, USA\\
$^{\dagger}$Idaho National Laboratory, 1955 Fremont Ave, Idaho Falls, ID 83415, USA\\
\{abinemra, rpaudyal, rupadhya, dwijesek\}@gmu.edu, arupjyoti.bhuyan@inl.gov
}
}

\maketitle
\begin{abstract}
The standard 5G Authentication and Key Agreement  protocol (5G-AKA) authenticates subscriber credentials but does not verify the integrity of user  equipment (UE). Existing literature includes admission-time device-certificate authentication and Trusted Platform Module (TPM)-based measured-boot attestation, which can additionally verify that the UE presents enrolled device credentials and boots into an approved state. However, these checks do not reveal whether harmful unauthorized software is executed after admission. We present TRACE-5G, which continuously verifies the UE's runtime software state and uses the resulting trust state to control network access. TRACE-5G retains the pre-admission certificate and boot checks, and in addition places the admitted UE in restricted connectivity while runtime integrity is verified. Linux Integrity Measurement Architecture (IMA) records runtime software measurements, TPM PCR (Platform Configuration
Register) cryptographically accumulates them, and Keylime, a remote-attestation framework, repeatedly evaluates TPM and IMA evidence against enrolled references. A Trusted result grants full access on the existing PDU session, but an Untrusted result withholds or removes full access and causes network deregistration. We implement TRACE-5G on an srsRAN/Open5GS testbed with a hardware TPM2.0 UE and core-side firewall enforcement. Across ten trusted trials, Level 2 access was established in 9--12s. Across ten untrusted trials, promotion was consistently withheld; in a representative trace, restrictive policy was applied within 5ms of the Untrusted decision and UE-context release completed within 1.670s. TRACE-5G extends UE trust beyond admission without any modification of the standardized 5G protocols.
\end{abstract}
\noindent

\section{Introduction}
\label{sec:introduction}

The 5G security architecture uses 5G Authentication and Key Agreement
protocol (5G-AKA) to authenticate a subscriber and derive keys that protect Non-Access Stratum (NAS) and Access Stratum signaling~\cite{3gpp33501}. However, subscriber authentication does not establish that the physical user equipment (UE) remains trustworthy. Examples that violate would a valid USIM may be used in an unauthorized platform, a legitimate UE may boot from modified software, or an admitted UE may later execute unauthorized software. Admission-time device attestation addresses part of this
problem. Battlefield 5G adds two independent device-certificate checks
and Trusted Platform Module (TPM)-based Platform Configuration Register
(PCR)~0--7 boot attestation before normal 5G registration~\cite{battlefield5g}. These checks bind network admission to an enrolled device and approved boot state, but the resulting trust decision remains point-in-time because PCRs~0--7 do not describe software changes that occur after admission. Consequently, a UE that later executes unauthorized software may continue to retain network access because its runtime integrity is not re-evaluated.

\begin{figure}[!t]
\raggedright
\includegraphics[width=0.99\columnwidth]{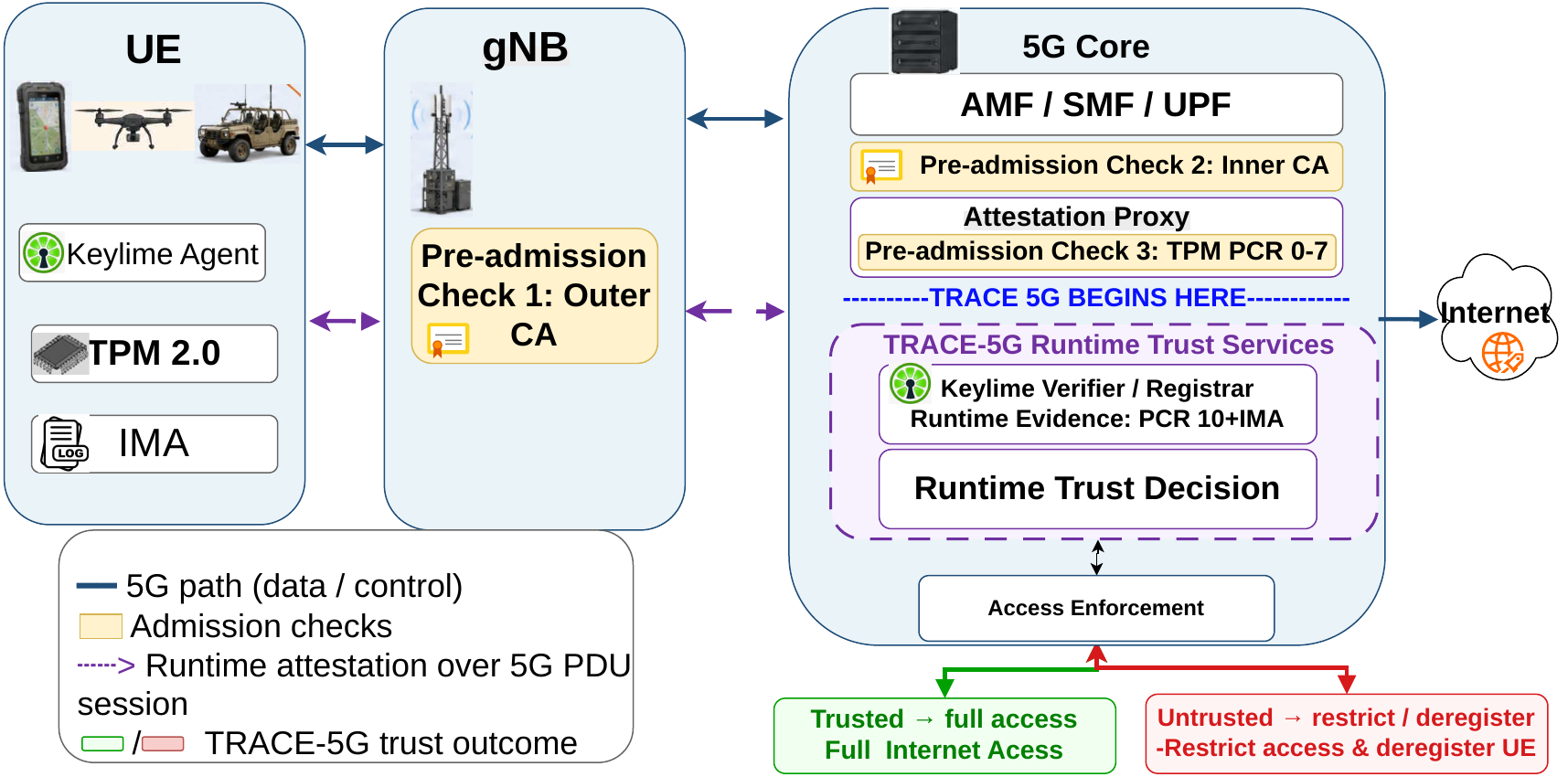}
\caption{TRACE-5G UE Trust and Enforcement Architecture}
\label{fig:architecture}
\end{figure}

Continuous attestation provides a way to evaluate integrity after this
initial decision. A TPM provides a hardware-backed root of trust and
protects attestation keys and measured state, while the Linux Integrity
Measurement Architecture (IMA) measures runtime files and executables \cite{berbecaru2023mitigating}.
Separately, Keylime is a remote-attestation framework that repeatedly evaluates TPM and IMA evidence to maintain an updated integrity state~\cite{schear2016keylime}. Emran et al. applied TPM, IMA, and Keylime with container-aware measurement attribution to Kubernetes-hosted 5G core functions~\cite{emran2025tpsisa}. That work attests 5G
infrastructure, whereas TRACE-5G addresses the continued authorization
of an active cellular UE whose runtime state can change after admission. Consequently, TRACE-5G treats network authorization as a continuously re-evaluated decision,
\(\mathcal{A}_{UE}(t)=f(I_{dev},I_{boot},I_{runtime}(t))\), where
\(I_{dev}\), \(I_{boot}\), and \(I_{runtime}(t)\) represent the enrolled
device identity, boot integrity, and runtime integrity at time \(t\).
A UE first satisfies the device and boot checks, enters restricted
connectivity, and receives full access only after its runtime state is
verified. Continued full access remains conditional on subsequent
runtime attestation. This approach aligns with Zero Trust Architecture (ZTA) principles by making continued network access conditional on current device posture rather than treating successful admission as permanent trust~\cite{dod2022ztra}.

Figure~\ref{fig:architecture} shows the TRACE-5G architecture. The
three pre-admission checks remain at their existing enforcement points:
Outer CA verification at the gNB, independent Inner CA verification at
the AMF, and TPM PCR~0--7 boot-attestation verification through the
core-side attestation proxy \cite{battlefield5g}. After these checks
succeed, standard 5G-AKA and registration proceed and the UE establishes a restricted PDU session. TRACE-5G then evaluates PCR~10 and container-aware IMA runtime evidence over the UE--gNB--core 5G path. A Trusted result grants full network access to the existing session, whereas an Untrusted result prevents or withdraws full-access
authorization and triggers network-requested deregistration. The contributions of this work are as follows.
\begin{itemize}

    \item We introduce a staged UE trust lifecycle that separates
    admission-time trust from runtime trust. After passing device-identity and boot-integrity checks, the UE remains restricted until its runtime software state is verified using TPM PCR~10 and IMA measurements. Full network access remains conditional on continued runtime trust verification.

    \item We connect the resulting Keylime trust state to 5G network access. A Trusted UE receives full access on its existing PDU session, while an Untrusted UE remains or becomes restricted and is subsequently deregistered. No  Data Network Name (DNN) change, address reassignment, or PDU-session re-establishment is required.

    \item We implement TRACE-5G on a hardware-TPM srsRAN/Open5GS testbed without introducing new standardized NAS/RRC message types. Across ten trusted trials, runtime trust and Level~2 access were established in approximately 9--12~s. Across ten untrusted trials, the same enforcement sequence was observed; in a representative trace, the restrictive policy followed the Untrusted decision within 5~ms and UE-context release is completed within 1.670~s.
\end{itemize}

\section{Background and Related Work}
\label{sec:relatedWork}

\subsection{5G-AKA and the Device-Trust Gap}
The 5G Authentication and Key Agreement (5G-AKA) protocol authenticates a subscriber and establishes the key hierarchy used to protect Non-Access Stratum (NAS) signaling between the UE and AMF and the Access Stratum signaling between the UE and gNB \cite{3gpp33501}. The procedure primarily authenticates the subscriber credential anchored in the USIM
rather than the integrity of the physical platform hosting that
credential. Formal analyses of 5G authentication similarly focus on
subscriber authentication, key agreement, unlinkability, and privacy, rather than the runtime software state of the UE~\cite{basin2018formal}.
Consequently, successful 5G-AKA does not by itself establish that the UE is an enrolled device, that it booted into an approved software state, or that its software remains trustworthy after authentication.

Prior cellular-security research has strengthened connection
bootstrapping and base-station authentication. Hussain et al. showed that weaknesses during cellular bootstrapping can enable multiple attacks \cite{hussain2019bootstrap}, while Singla et al. and other work proposed stronger mechanisms for secure 5G bootstrapping and base-station authentication~\cite{singla2021bootstrap,lotto2023baron,bhuyan2024real}.
These approaches improve the establishment of a secure cellular connection but they do not continuously evaluate the integrity of an already-connected UE. Emran et al. proposed Battlefield 5G to address the complementary
pre-admission device-trust problem~\cite{battlefield5g}. Their work introduces
three checks performed before normal registration proceeds: (1) an Outer CA certificate verification at the gNB, (2) an independent Inner CA verification at the AMF, and (3) TPM-based boot attestation using PCRs~0--7. These checks allow a credential-validation, but unauthorized or boot-tampered devices to be rejected before admission. However, the resulting trust decision is established only at connection time; The boot evidence from PCR 0--7 does not describe software changes that occur later
during the active UE session.

\subsection{TPM, IMA, and Remote Attestation}
\label{ssec:tpm-IMA}

A Trusted Platform Module (TPM)  is a hardware security component that provides protected cryptographic operations, key storage, and Platform Configuration Registers (PCRs) that support hardware-rooted integrity measurement~\cite{tcgtpm20}. During measured boot, measurements of firmware, bootloader, kernel, and other component of the platform are cumulatively extended to selected PCRs. Then an Attestation Key (AK) can sign a fresh TPM quote over those PCR values and a verifier-provided nonce, allowing a remote verifier to check both platform state and evidence freshness.

Linux Integrity Measurement Architecture (IMA) extends integrity measurement to runtime software. IMA records measurements of files and executables according to a configured policy and extends their digests into PCR~10. During remote attestation, the verifier checks the AK-signed quote and nonce, verifies that the IMA measurement history is consistent with the quoted PCR~10 value, and compares measured objects with configured reference values or allow-lists. This follows the general Internet Engineering
Task Force (IETF) Remote ATtestation ProcedureS (RATS) model, in which an attester provides evidence that a verifier evaluates against trusted reference values to produce an attestation result~\cite{birkholz2023rats}.

\subsection{Keylime Framework}
\label{ssec:KeyLime}

Keylime is an open-source framework for TPM and IMA-based remote attestation~\cite{schear2016keylime}. A Registrar maintains TPM identity information, an Agent on the attested platform provides TPM and IMA evidence, and a Verifier repeatedly evaluates that evidence to maintain an updated trust state. Operational studies have demonstrated Keylime's ability to detect measured runtime integrity violations while also highlighting practical issues such as allow-list maintenance and measurement coverage~\cite{ruffin2025keylime}. Emran et al. applied Keylime with container-aware IMA attribution to continuously attest Kubernetes-hosted 5G core functions~\cite{emran2025tpsisa}. In that work, the attested subjects were 5G Core components. Instead TRACE-5G  attests an active UE and connects its runtime trust state to continued network authorization. Table~\ref{tab:positioning} summarizes the difference between standard 5G-AKA, Battlefield 5G, and TRACE-5G in terms of the coverage of UE trust and the enforcement of the network.
\begin{table}[!t]
\caption{Positioning of TRACE-5G in the UE Trust Lifecycle}
\label{tab:positioning}
\centering
\scriptsize
\setlength{\tabcolsep}{1.8pt}
\renewcommand{\arraystretch}{1.08}

\begin{tabular}{
|>{\raggedright\arraybackslash}p{1.05cm}
|>{\raggedright\arraybackslash}p{2.87cm}
|>{\raggedright\arraybackslash}p{1.35cm}
|>{\raggedright\arraybackslash}p{2.50cm}|}
\hline

\textbf{Approach} &
\textbf{UE Trust Checks} &
\textbf{UE Lifecycle Coverage} &
\textbf{Access / Enforcement Outcome} \\
\hline

5G-AKA \cite{3gpp33501} &
Subscriber authentication using USIM credentials &
Authentication &
Authenticates the subscriber and establishes security keys \\
\hline

Battlefield 5G \cite{battlefield5g} &
Check 1: Outer certificate at the gNB;  \newline Check 2: Inner certificate at the AMF;  \newline Check 3: TPM PCR 0--7 boot attestation verified by the core attestation proxy &
Pre-admission only &
Admits UE if all three checks succeed; else reject\\
\hline

\textbf{TRACE-5G} &
\textbf{Perform Checks 1--3 for pre-admission assurance; \newline 
Add continuous post-admission UE runtime attestation using TPM PCR 10 and container-aware IMA measurements, verified by the core-side Keylime Verifier} &
\textbf{Pre-admission checks followed by continuous post-admission runtime verification} &
\textbf{Post-admission, keeps the UE in restricted access until runtime trust is established; then grants full access. \newline If runtime trust is lost, revoke access deregister the UE} \\
\hline

\end{tabular}
\end{table}

\section{Threat Model and Design Goals}
\label{sec:threat}

We consider a private/tactical 5G deployment where the operator controls the gNB, core, Keylime services, enrollment infrastructure, and TPM trust anchors. An adversary may transplant a valid SIM, use an unauthorized device, modify boot components, introduce unauthorized runtime software/containers, modify measured executable content, or replay captured evidence. We assume that the attacker cannot extract TPM-resident keys, forge trusted-CA signatures or fresh TPM quotes, or subvert the trusted kernel/IMA path. RF denial-of-service, TPM compromise, and attacks outside the configured IMA measurement coverage are beyond the scope. Our design goals are: \emph{G1} verify enrolled identity and boot state before admission; \emph{G2} bootstrap attestation with restricted connectivity; \emph{G3} repeatedly verify runtime integrity; \emph{G4} translate trust loss into data- and control-plane enforcement; and \emph{G5} preserve standardized NAS/RRC message formats.

\section{TRACE-5G System Design}
\label{sec:design}

\begin{figure}[!t]
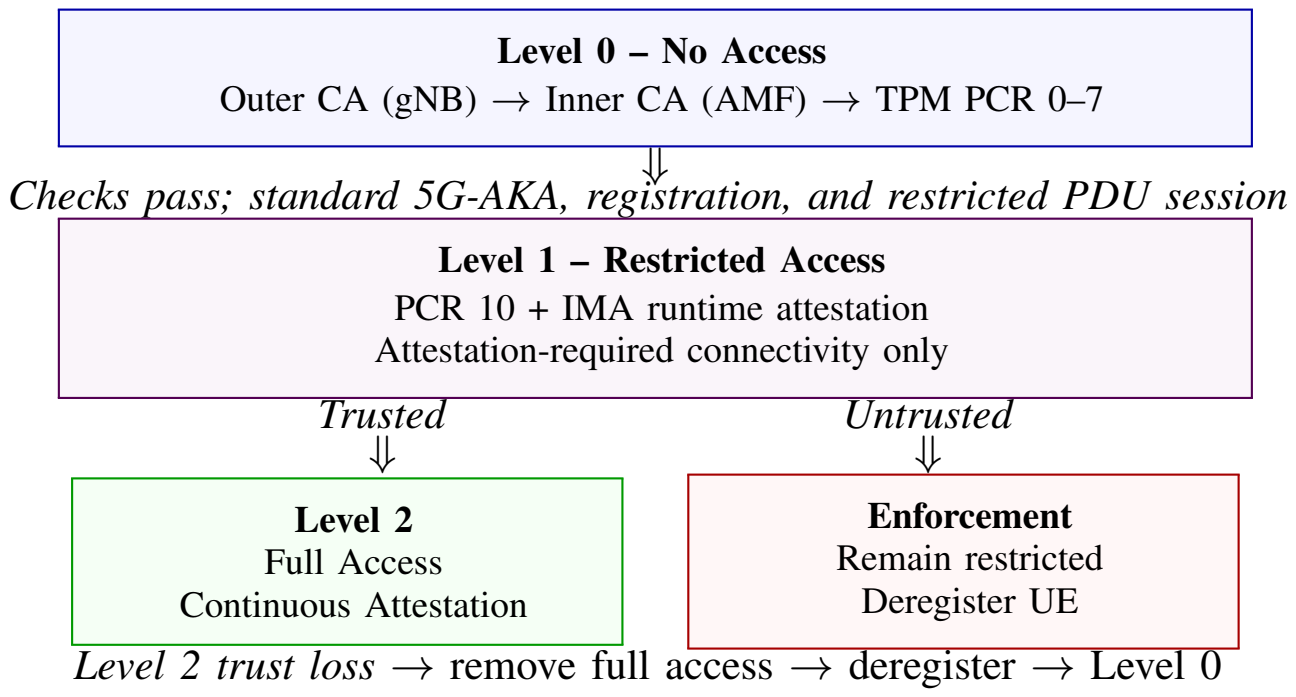

\centering
\scriptsize
\setlength{\fboxsep}{5pt}

\fcolorbox{blue!65!black}{blue!4}{
\parbox{0.84\columnwidth}{\centering
\textbf{Level 0 -- No Access}\\[1pt]
Outer CA (gNB) $\rightarrow$ Inner CA (AMF) $\rightarrow$ TPM PCR~0--7
}}

{\footnotesize
$\Downarrow$\\[-2pt]
\textit{Checks pass; standard 5G-AKA, registration, and restricted PDU session}
}

\fcolorbox{violet!65!black}{violet!4}{
\parbox{0.84\columnwidth}{\centering
\textbf{Level 1 -- Restricted Access}\\[1pt]
PCR~10 + IMA runtime attestation\\
Attestation-required connectivity only
}}

\begin{tabular}{c@{\hspace{12pt}}c}
\parbox{0.36\columnwidth}{\centering
{\footnotesize
\textit{Trusted}\\[-1pt]
$\Downarrow$
}}
&
\parbox{0.36\columnwidth}{\centering
{\footnotesize
\textit{Untrusted}\\[-1pt]
$\Downarrow$
}}
\end{tabular}

\begin{tabular}{c@{\hspace{12pt}}c}

\fcolorbox{green!60!black}{green!4}{
\parbox{0.36\columnwidth}{\centering
\textbf{Level 2}\\
Full Access\\
Continuous Attestation
}}
&
\fcolorbox{red!65!black}{red!3}{
\parbox{0.36\columnwidth}{\centering
\textbf{Enforcement}\\
Remain restricted\\
Deregister UE
}}
\end{tabular}

{\footnotesize
\textit{Level 2 trust loss}
$\rightarrow$
remove full access
$\rightarrow$
deregister
$\rightarrow$
Level 0
}

\caption{TRACE-5G UE trust lifecycle.}
\label{fig:lifecycle}
\end{figure}

TRACE-5G organizes UE trust into three levels. \emph{Level~0} represents pre-admission device and boot assurance with no network access; \emph{Level~1} provides restricted connectivity while runtime trust is established; and \emph{Level~2} provides full connectivity conditioned on continuous attestation. An untrusted verdict prevents promotion to Level~2 or withdraws existing Level~2 authorization and returns the UE to Level~0 by deregistration. Figure~\ref{fig:lifecycle} summarizes this lifecycle. TRACE-5G also preserves the standard 5G control-plane procedures. After the Checks~1--3 are successful, normal 5G-AKA and registration are executed, and Level~1 connectivity is established through the standard establishment of the PDU session. Promotion to Level~2 is only a core-side policy update on the existing session, while trust loss uses standard network-requested deregistration and UE-context release. No new NAS or RRC message types or message formats are introduced.

\subsection{One-Time Enrollment}
\label{ssec:1timeEnrollment}

Before deployment, the UE is enrolled in a trusted environment. The activation of the TPM endorsement key (EK) and the attestation key (AK) binds the AK to the intended hardware TPM, while PCR~0--7 from a known-good boot establish the boot reference values and PCR~10 is included in the runtime-attestation policy. The operator generates host and per-container allow-lists, and TRACE-5G uses the \texttt{ima-dep-cgn}~\cite{ferro2026privacy} template with the \texttt{cgpath} field to identify whether each measurement belongs to the host or a specific container. The UE's Keylime Agent is registered with the attestation infrastructure and associated with a persistent Keylime identity. A core-side enrollment database stores the EK/AK metadata, PCR policy, Keylime identity, runtime reference information, and session-association metadata required for subsequent admission and runtime verification. These trusted values and allow-lists remain at the core and are not supplied by the fielded UE during attestation.

\subsection{Level 0: Admission-Time Assurance}
\label{ssec:l0}

TRACE-5G uses as its Level~0 foundation the three-stage pre-authentication design introduced by~\cite{battlefield5g}. After RRC establishment, the gNB performs Check~1 by challenging the UE and validating its Outer CA device certificate and nonce signature before forwarding the Registration Request. At AMF, Check~2 independently validates the Inner CA certificate and a fresh nonce signature. Check~3 then challenges the hardware TPM for an AK-signed quote over PCRs~0--7; the core-side attestation proxy verifies the AK, quote signature, freshness, and PCR values against the enrolled boot baseline. The Registration Request is preserved while these checks execute and enters normal processing only after all three succeed. Then the standard 5G-AKA and registration proceed. Checks~1--3 establish device identity and boot integrity at admission time, but do not describe the subsequent runtime software state; TRACE-5G therefore places the admitted UE into Level~1 rather than granting unrestricted access immediately.

\subsection{Level~1: Restricted Access}
\label{ssec:l2}

After Level~0 succeeds, TRACE-5G initializes runtime monitoring and allows standard 5G registration to complete. The UE then establishes a PDU session through a dedicated restricted DNN using the standard PDU Session Establishment procedure. This places the UE in Level~1. The core-side access policy permits only the connectivity required for attestation and withholds normal full-access privileges until runtime trust is established. A persistent Keylime Agent runs on the UE and communicates with the core-side attestation infrastructure through this restricted 5G PDU session. Thus, runtime attestation is performed over the same radio-connected data path whose access is being evaluated.

For each attestation cycle, the Agent provides a fresh TPM quote over PCR~10 together with the IMA measurement information. The Keylime Verifier validates the quote signature and freshness, checks that the IMA measurement history is consistent with the quoted PCR~10 value, and compares the measured hashes with the enrolled allow-lists. Using the container-aware \texttt{cgpath} field, host measurements are evaluated against the host allowlist, while container measurements are evaluated against the corresponding container-specific allow-list. The resulting verdict represents the UE's current runtime trust. An unexpected executable, unauthorized container, or measured object whose digest differs from its approved reference value can produce an untrusted verdict. In that case, Level~2 promotion is withheld, the restricted policy remains active, and the UE proceeds towards deregistration. Only a UE whose runtime measurements satisfy the configured policy advances to Level~2. TRACE-5G therefore separates successful network admission from authorization for full network use.

\subsection{Level 2: Full Access with Continuous Monitoring}
\label{ssec:l2ContinuousMonitoring}

When the Keylime Verifier reports that the UE is trusted, TRACE-5G promotes the UE from restricted connectivity at Level~1 to full access at Level~2. The attestation proxy updates the core-side access policy for the existing session without any DNN change, IP-address reassignment, or PDU-session re-establishment. Promotion does not terminate attestation. The Keylime Verifier continuously obtains fresh PCR~10 and IMA evidence and updates the runtime trust state of UE. Level~2 therefore represents full access conditioned on continued compliance with the configured runtime integrity policy rather than a permanent trust decision.

\subsection{Revocation and Re-Admission} 
\label{ssec:revocation}

If continuous runtime attestation reports that the UE is untrusted while in Level~2, TRACE-5G removes the full-access firewall policy and reverts back to the restrictive policy for the existing session. The AMF then observes the untrusted state and initiates standard network-requested deregistration, followed by UE-context release. If the UE is still in Level~1 when it is reported as untrusted, the restrictive firewall policy is already in place. Therefore, TRACE-5G withholds the promotion of Level~2, and AMF proceeds with deregistration and release in the context of UE. After deregistration, the previous runtime-trust state is cleared. If the UE reconnects, it must restart from Level~0 and again pass Check~1, Check~2, and Check~3 before standard 5G registration proceeds and the UE can re-enter Level~1 runtime-trust establishment. Consequently, TRACE-5G connects runtime attestation directly to network authorization through \emph{measure $\rightarrow$ attest $\rightarrow$ verify $\rightarrow$ authorize $\rightarrow$ enforce}.

\section{Implementation and Experimental Setup}
\label{sec:implementation}

\begin{figure}[!t]
\raggedleft
\includegraphics[width=0.98\columnwidth]{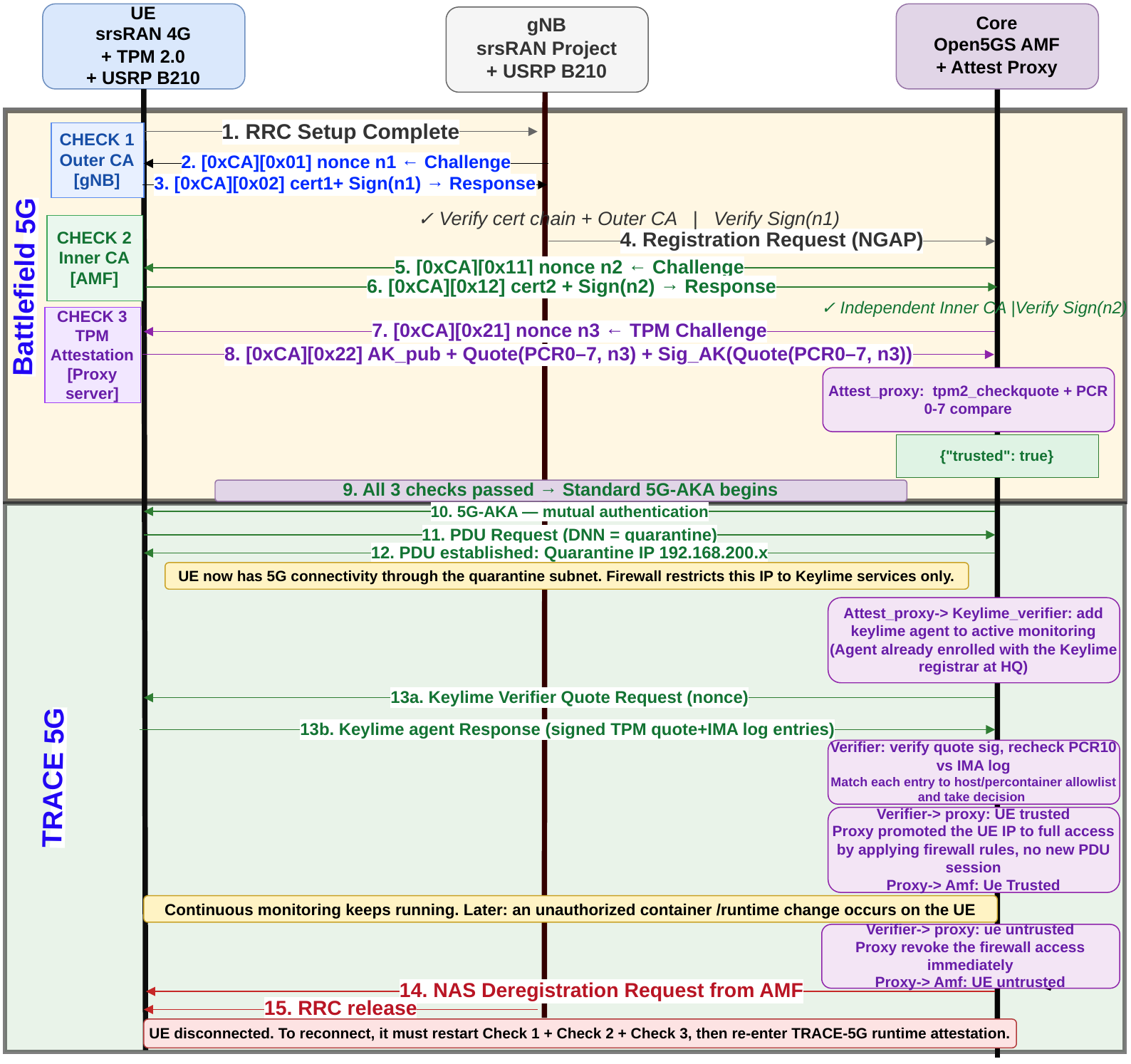}
\caption{End-to-end TRACE-5G communication and enforcement flow.}
\label{fig:traceflow}
\end{figure}

We implemented TRACE-5G on a three-machine 5G SA testbed. The core
runs Open5GS in Docker, with the Keylime Verifier and Registrar on the host and the attestation proxy in Docker. The gNB uses srsRAN Project with an Ettus USRP B210. The UE uses srsUE in 5G SA mode with an Ettus USRP B200, a Nuvoton NPCT75x hardware TPM~2.0, and a custom Linux 5.13.19 kernel containing the container-aware IMA template. Figure~\ref{fig:traceflow} shows the communication flow from pre-admission verification to runtime attestation and enforcement.

\subsection{Enrollment and Pre-Admission Integration}
\label{ssec:enrollment+Integration}

For the prototype, the one-time enrollment described in Section~\ref{sec:design} was completed prior to the experiments. The UE's Keylime Agent was registered with the core-side Keylime Registrar, while the enrollment database and trusted reference information were maintained on the core. The enrollment record associated the UE with its TPM and Keylime identities, PCR policy, runtime reference information, and restricted-session information used during attestation and enforcement.

The pre-admission implementation retains the three-check mechanism of Battlefield 5G~\cite{battlefield5g}. For Check~1, the gNB RRC path holds the UE's Registration Request while the Outer CA certificate and
nonce signature are verified. The request is forwarded to the AMF
through NGAP only after Check~1 succeeds. At the AMF, the Registration Request is temporarily saved while Checks~2 and~3 are executed. The Check~2 validates the independent Inner CA certificate and the nonce signature. For Check~3, the AMF forwards the UE's TPM evidence to the attestation proxy. The proxy verifies the enrolled AK, quote signature, challenge freshness, and PCR~0--7 values against the enrolled boot baseline. After Check~3 succeeds, the AMF creates the per-UE TRACE-5G runtime state and starts the runtime-monitoring process. The saved Registration Request is then returned to the normal Open5GS path, where standard 5G-AKA and registration continue. Runtime trust remains pending until the UE obtains the connectivity required for Keylime attestation.

\subsection{Runtime Attestation and Enforcement Implementation}
\label{ssec:runtime}

After registration, the UE establishes a PDU session through the
dedicated restricted DNN. The Open5GS SMF/UPF provides the UE with its restricted-session address, which is associated with the corresponding UE and Keylime runtime state. The core-side firewall policy initially permits the connectivity required for runtime attestation while
withholding normal full-access traffic. The UE runs a persistent Keylime Agent with the hardware TPM and the container-aware IMA kernel. Once the restricted PDU session is available, the Agent communicates with the core-side Keylime infrastructure through the UE's 5G PDU-session interface.

The UE IMA policy measures executable launches using
\texttt{BPRM\_CHECK} with \texttt{MAY\_EXEC} and executable memory
mappings using \texttt{FILE\_MMAP} with \texttt{MAY\_EXEC}. The
container-aware IMA template includes the \texttt{cgpath} field, which identifies the execution context of each measurement. Host measurements are compared with the enrolled host allow-list, while measurements from
a container are compared with the corresponding container-specific
allow-list. The Keylime Verifier evaluates these IMA measurements
together with fresh TPM evidence over PCR~10 and maintains the UE's
runtime trusted state. The Open5GS AMF was extended with per-UE Keylime state and functions to start, poll, and stop runtime monitoring. The attestation proxy associates the UE identity, Keylime identity, and restricted-session information with the trust state reported by the Verifier. The proxy then maps that trust state to the core-side firewall policy. A trusted
result grants full access to the existing PDU session without changing the DNN, UE address, or PDU session. If the UE is reported as untrusted, the restrictive firewall policy remains in place or is restored if full access had already been granted.

The Keylime Verifier performs runtime attestation every 2~seconds. The attestation proxy queries the Verifier every 3~s, while the AMF queries the proxy every 5~s. The proxy acts directly on the Verifier state when updating the firewall policy. Therefore, access enforcement does not wait for the AMF polling cycle. The AMF independently observes the untrusted state through its poll and initiates standard network-requested deregistration. Runtime-monitoring state is stopped and cleared when the UE context is released. Verification remains pending while the configured retry process is active. Keylime runtime attestation and trust-dependent network enforcement were exercised through the UE's restricted 5G PDU-session path.

\section{Evaluation}
\label{sec:evaluation}

The evaluation focuses on the functionality introduced by TRACE-5G:
post-admission runtime-trust establishment, continuous attestation over the UE's 5G PDU-session path, trust-driven access promotion, and enforcement of an untrusted runtime state. The Level~0 mechanisms are retained from Battlefield 5G and are not re-evaluated as new contributions. Emran et al. previously demonstrated rejection of SIM transplantation, invalid device credentials, boot tampering, and replay, and reported 373.4~ms of average added onboarding overhead across six
trials~\cite{battlefield5g}. In TRACE-5G, Checks~1--3 were exercised as part of the complete admission-to-runtime path.  All runtime experiments used the attestation
configuration described in Section~\ref{sec:implementation}.

\begin{table}[!t]
\caption{Security Coverage of Battlefield 5G and TRACE-5G}
\label{tab:securitycoverage}
\centering
\scriptsize
\setlength{\tabcolsep}{2pt}
\renewcommand{\arraystretch}{1.0}

\begin{tabular*}{\columnwidth}{
@{\extracolsep{\fill}}
|>{\raggedright\arraybackslash}p{0.23\columnwidth}
|>{\raggedright\arraybackslash}p{0.27\columnwidth}
|>{\raggedright\arraybackslash}p{0.38\columnwidth}|
}
\hline

\textbf{Condition} &
\textbf{Battlefield 5G} &
\textbf{TRACE-5G} \\
\hline

Unauthorized UE / invalid credential &
Blocked by Outer and Inner certificate checks &
Same admission protection retained \\
\hline

Boot tampering / replay &
PCR~0--7 and fresh signed evidence &
Same boot and replay protection retained \\
\hline

Runtime software / container violation &
Outside admission scope &
Detected through PCR~10/IMA and enrolled allowlists \\
\hline

Untrusted before Level~2 &
No runtime trust stage &
Promotion withheld; UE remains restricted and is deregistered \\
\hline

Trust loss after Level~2 &
No continuous enforcement &
Full access withdrawn; restrictive policy restored; UE deregistered \\
\hline

\end{tabular*}
\end{table}

\subsection{End-to-End Trust-Lifecycle Validation}
\label{ssec:validation}

The complete TRACE-5G path was exercised with an enrolled,
policy-compliant UE. The UE first passed Checks~1--3, completed standard 5G-AKA and registration, and established the restricted PDU session. Using this restricted PDU session, the UE's Keylime Agent exchanged attestation evidence with the core-side Keylime infrastructure. The Keylime Verifier then evaluated a fresh TPM quote over PCR~10 together with the container-aware IMA measurements. When the measurements satisfied the enrolled host and container policies, the UE was reported as Trusted and the core-side access policy was updated to provide Level~2 full access. The DNN, UE address, and PDU session remained unchanged, and continuous attestation continued after promotion.

The untrusted path was also repeated ten times. A container-level
allow-list violation was already present when runtime-trust establishment began. Therefore the UE passed the Level~0 identity and boot checks and completed registration, but its runtime measurements did not satisfy the enrolled policy. The trust state moved from \emph{Pending} to \emph{Untrusted} while the UE was still in Level~1, so full-access promotion was withheld. The restrictive firewall policy remained active, the AMF subsequently initiated network-requested deregistration, and the UE context was released. The same enforcement sequence and comparable response timing were observed across the ten repetitions. Post-Level~2 trust-loss revocation was also functionally exercised by
introducing an unauthorized container after full access had been granted,
which caused the UE to become Untrusted and lose full access. As observed, these experiments validate the three runtime outcomes of TRACE-5G: Pending remains restricted, Trusted receives Level~2 access, and Untrusted is prevented from obtaining full network authorization. Table~\ref{tab:securitycoverage} summarizes the resulting security coverage.

\begin{table}[!t]
\caption{TRACE-5G Runtime Trust and Enforcement Measurements}
\label{tab:timing}
\centering
\scriptsize
\setlength{\tabcolsep}{1.0pt}
\renewcommand{\arraystretch}{1.08}

\begin{tabular}{
|>{\raggedright\arraybackslash}p{4cm}
|>{\centering\arraybackslash}p{1.25cm}
|>{\raggedright\arraybackslash}p{2.55cm}|}
\hline

\textbf{Measured Interval} &
\textbf{Result} &
\textbf{Interpretation} \\
\hline

Runtime monitoring start $\rightarrow$ Verifier Trusted visible &
$\sim$6.3~s &
Runtime trust becomes visible \\
\hline

Verifier Trusted visible $\rightarrow$ full-access policy activation &
$\sim$3.0~s &
Proxy propagation and policy update \\
\hline

Level~2 full-access activation $\rightarrow$ AMF observes Trusted &
$\sim$0.7~s &
Subsequent AMF observation \\
\hline

\textbf{Level~1 runtime-trust establishment $\rightarrow$ Level~2 full-access activation} &
\textbf{$\sim$9--12~s} &
\textbf{Post-admission restricted-to-full-access promotion} \\
\hline

Untrusted decision $\rightarrow$ restrictive policy applied &
5~ms &
Immediate policy enforcement \\
\hline

Restrictive policy applied $\rightarrow$ AMF observes Untrusted &
677~ms &
Trust-state propagation to AMF \\
\hline

AMF observes Untrusted $\rightarrow$ NAS Deregistration sent &
308~ms &
Control-plane reaction \\
\hline

NAS Deregistration sent $\rightarrow$ UE-context release &
680~ms &
Connection teardown \\
\hline

\textbf{Untrusted decision $\rightarrow$ UE-context release} &
\textbf{1.670~s} &
\textbf{Post-verdict complete disconnection} \\
\hline

\end{tabular}
\end{table}

\subsection{Runtime Trust and Enforcement Timing}
\label{ssec:runtimeTrust}

Table~\ref{tab:timing} summarizes the timing results for the Trusted and Untrusted cases. Across ten trusted-promotion trials, TRACE-5G reached Level~2 full-access policy activation approximately 9--12s after runtime monitoring began. This value represents the complete runtime-trust establishment interval rather than the processing time of a single TPM
quote or the delay of the firewall action itself. In a representative trace, approximately 6.3s elapsed before the Verifier exposed the Trusted state. This period includes the initial Keylime runtime verification and the asynchronous attestation process. After the Trusted state became available, full-access policy activation followed approximately 3s later, which is consistent with the proxy's 3s Verifier-polling interval. Thus, most of the observed 9--12~s delay is associated with establishing and propagating the initial runtime-trust
result, while the subsequent authorization change requires
approximately one proxy polling cycle.

In the untrusted experiments, an additional container not covered by the enrolled container allow-lists was deployed on the UE before runtime attestation began. When Keylime evaluated the UE's PCR~10 and IMA evidence, measurements from this container caused the UE to be reported as Untrusted. Once the Untrusted verdict became available, the restrictive firewall policy was applied within 5~ms. The AMF observed the Untrusted state 677ms after the policy action and issued the network-requested NAS Deregistration Request after a further 308ms.
UE-context release completed 680ms later, giving 1.670s from the
Untrusted decision to complete UE-context release. Because the
additional container was deployed before runtime attestation began,
these measurements do not represent container-deployment-to-detection latency. These measurements show that initial runtime-trust establishment
requires several seconds, while post-verdict enforcement occurs much
faster.

\section{Conclusion}
\label{sec:conclusion}

TRACE-5G extends the trust of UE beyond admission by combining device and boot assurance with continuous runtime attestation and trust-driven network enforcement. The prototype keeps an admitted UE restricted until its runtime state is verified, then grants full access only while that trust is maintained. Across ten trusted trials, Level~2 access was established
in approximately 9--12~s, representing the time required to
establish runtime assurance before unrestricted access. This is a
security--latency tradeoff for tactical and private 5G environments, where verifying the current integrity of the UE can be more important than granting immediate full access. Once an Untrusted decision is available, restrictive policy is applied within 5~ms in a representative trace and UE-context release completes within 1.670~s. TRACE-5G provides this protection while preserving standard 5G registration, PDU-session, and deregistration procedures.

\bibliography{custom}
\bibliographystyle{IEEEtran}

\end{document}